\documentclass[runningheads]{llncs}

\usepackage[T1]{fontenc}

\usepackage{graphicx}

\usepackage{caption}
\usepackage{booktabs}
\usepackage{float}
\usepackage{flushend}
\usepackage{multirow} 
\usepackage{threeparttable} 
\usepackage[nolist]{acronym}
\usepackage{cite}
\usepackage{amsmath,amssymb,amsfonts}
\usepackage{algorithmic}
\usepackage{textcomp}
\usepackage{xcolor}
\usepackage{subfigure}
\usepackage[pscoord]{eso-pic}
\newcommand{\placetextbox}[3]{
  \setbox0=\hbox{#3}
  \AddToShipoutPictureFG*{
    \put(\LenToUnit{#1\paperwidth},\LenToUnit{#2\paperheight}){\vtop{{\null}\makebox[0pt][c]{#3}}}%
  }%
}%

\begin{document}
\title{ONNX-to-Hardware Design Flow for Adaptive Neural-Network Inference on FPGAs}

\author{Federico Manca \inst{1} \and
Francesco Ratto \inst{1}\orcidID{0000-0001-5756-5879} \and
Francesca Palumbo \inst{2}\orcidID{0000-0002-6155-1979}}

\authorrunning{F. Manca et al.}
\titlerunning{ONNX-to-HW Design Flow.}

\institute{University of Sassari,  Piazza Università 21, 07100 Sassari, Italy\\
\email{\{fmanca2, fratto\}@uniss.it}\\ \and
University of Cagliari, via Marengo 2, 09123 Cagliari, Italy\\
\email{francesca.palumbo@unica.it}}

\maketitle



\begin{acronym}
\acro{ac}[AC]{Approximate Computing}
\acro{aes}[AES]{Advanced Encryption Standard}
\acro{api}[API]{Application Programming Interface}
\acrodefplural{api}[APIs]{Application Programming Interfaces }
\acro{asic}[ASIC]{Application Specific Integrated Circuit}

\acro{bnn}[BNN]{Binary Neural Network}
\acrodefplural{bnn}[BNNs]{Binary Neural Networks}

\acro{cgr}[CGR]{Coarse-Grain Reconfigurable}
\acro{cpg}[CPG]{Co-Processor Generator}
\acro{cps}[CPS]{Cyber-Physical System}
\acrodefplural{cps}[CPSs]{Cyber-Physical Systems}
\acro{cpu}[CPU]{Central Processing Unit}
\acrodefplural{cpu}[CPUs]{Central Processing Units}
\acro{cnn}[CNN]{Convolutional Neural Network}

\acro{dag}[DAG]{Directed Acyclic Graph}
\acro{dnn}[DNN]{Deep Neural Network}
\acro{dpn}[DPN]{Dataflow Process Network}
\acrodefplural{dpn}[DPNs]{Dataflow Process Networks}
\acro{dse}[DSE]{Design Space Exploration}
\acro{dsp}[DSP]{Digital Signal Processing}


\acro{fft}[FFT]{Fast Fourier Transform}
\acro{fifo}[FIFO]{First-In First-Out queue}
\acrodefplural{fifo}[FIFOs]{First-In First-Out queues}
\acro{fpga}[FPGA]{Field Programmable Gate Array}
\acrodefplural{fpga}[FPGAs]{Field Programmable Gate Arrays}

\acro{gpu}[GPU]{Graphics Processing Unit}


\acro{hevc}[HEVC]{High Efficiency Video Coding}
\acro{hls}[HLS]{High Level Synthesis}
\acro{hw}[hw]{hardware}
\acro{hwpu}[HWPU]{HW Processing Unit}
\acrodefplural{hwpu}[HWPUs]{HW Processing Unit (HWPU)s}

\acro{ip}[IP]{Intellectual Property}
\acrodefplural{ip}[IPs]{Intellectual Properties}


\acro{lut}[LUT]{Look-Up Table}
\acrodefplural{lut}[LUTs]{Look-Up Tables}

\acro{mdc}[MDC]{Multi-Dataflow Composer}
\acro{mdg}[MDG]{Multi-Dataflow Generator}
\acro{mlp}[MLP]{Multi Layer Perceptron}
\acro{mcdma}[MCDMA]{Multi-channel DMA}
\acro{moa}[MoA]{Model of Architecture}
\acrodefplural{moa}[MoAs]{Models of Architecture}
\acro{moc}[MoC]{Model of Computation}
\acrodefplural{moc}[MoCs]{Models of Computation}
\acro{mpsoc}[MPSoC]{Multi-Processor \ac{soc}}
\acrodefplural{mpsoc}[MPSoCs]{Multi-Processor System on Chip}

\acro{nn}[NN]{Neural Network}

\acro{os}[OS]{Operating System}

\acro{pc}[PC]{Platform Composer}
\acro{pe}[PE]{Processing Element}
\acro{pisdf}[PiSDF]{Parameterized and Interfaced Synchronous DataFlow}
\acro{pdf}[PDF]{Parameterized DataFlow}
\acro{pl}[PL]{Programmable Logic}
\acro{ps}[PS]{Processing System}
\acro{pmc}[PMC]{Performance Monitoring Counter}
\acro{psdf}[PSDF]{Parameterized Synchronous DataFlow}

\acro{qsoc}[QSoC]{Quantized \ac{soc}}
\acro{qat}[QAT]{Quantization-aware Training}

\acro{qnn}[QNN]{Quantized Neural Network}

\acro{rtl}[RTL]{Register Transfer Level}

\acro{sg}[SG]{Scatter-Gather}
\acro{sdf}[SDF]{Synchronous DataFlow}
\acro{soc}[SoC]{System on a Chip}
\acro{smmu}[SMMU]{System Memory Management Unit}

\acro{til}[TIL]{Template Interface Layer}

\acro{uav}[UAV]{Unmanned Aerial Vehicle}
\acro{ugv}[UGV]{Unmanned Ground Vehicle}





\end{acronym}

\placetextbox{0.5}{1}{\texttt{\small This is the authors' version of the work. It is posted here for your personal use. Not for redistribution.}}
\placetextbox{0.5}{0.98}{\texttt{\small The definitive Version of Record will appear in }}%
\placetextbox{0.5}{0.96}{\texttt{\textit{\small Proceedings of the 24th International Conference, SAMOS 2024, Samos, Greece, June 29 - July 4, 2024.}}}
\placetextbox{0.5}{0.94}{\texttt{\small Please refer to the reviewed and published Version of Record for citing this work. }}

\begin{abstract}
\acp{nn} provide a solid and reliable way of executing different types of applications, ranging from speech recognition to medical diagnosis, speeding up onerous and long workloads. 
The challenges involved in their implementation at the edge include providing diversity, flexibility, and sustainability. That implies, for instance,  supporting evolving applications and algorithms energy-efficiently.
Using \ac{hw} or software accelerators can deliver fast and efficient computation of the \acp{nn}, while flexibility can be exploited to support long-term adaptivity.
Nonetheless, handcrafting a \ac{nn} for a specific device, despite the possibility of leading to an optimal solution, takes time and experience, and that's why frameworks for \ac{hw} accelerators are being developed. 
This work, starting from a preliminary semi-integrated ONNX-to-hardware toolchain~\cite{ratto2023}, focuses on enabling \ac{ac} leveraging the distinctive ability of the original toolchain to favor adaptivity.
The goal is to allow lightweight adaptable \ac{nn} inference on FPGAs at the edge. 

\end{abstract}

\section{Introduction} \label{sec::intro}
\ac{cps} integrate \textit{``computation with physical processes whose behavior is deﬁned by both the computational (digital and other forms) and the physical parts of the system''}\footnote{https://csrc.nist.gov/glossary/term/cyber physical systems}. These systems are characterized by significant information exchange with the environment and dynamic, reactive behaviors in response to environmental changes.
In modern systems, whether \ac{cps} or not, NN-assisted decision-making can be directly deployed at the edge on small embedded platforms. This approach reduces latency, energy consumption, and often ensures higher privacy levels \cite{edge_ai}. Nonetheless, executing AI models on resource-constrained edge devices presents several challenges, including limited computing and memory capacities. Balancing model accuracy and execution efficiency exposes a crucial design trade-off.

In response to these challenges, \acp{fpga} devices emerge as a valuable choice for NN inference at the edge ~\cite{guo2019}. They can guarantee \ac{hw} acceleration, execution flexibility, and energy efficiency thanks to the possibility of tailoring the \ac{hw} architecture to the specific application. 
Despite existing solutions, there remains a lack of full support for advanced features, particularly the adaptivity naturally supported over these kind of platforms. Computing adaptivity empowers CPS to thrive in complex, ever-changing environments.

This paper aims to take steps towards filling that gap. The goal is featuring adaptivity targeting \ac{cnn} models as applications and edge \acp{fpga} as computing platforms. \ac{cnn} models have proven to be positively affected by the application of \ac{ac} methodologies \cite{ac_survey}. State of the art approaches \cite{aarestad2021,fraser2017} apply it in a data-oriented manner, as discussed in Sect. \ref{ssec::streaming-archs}. In this paper, the combination of data-oriented and computation-oriented strategies targeting runtime adaptivity, by means of reconfiguration, is presented. Different execution profiles are operated at runtime by an adaptive inference engine. This latter is developed with the proposed design flow. In summary, the contributions of this work are:
\begin{itemize}
    \item A novel design flow that enables the inference of Quantized ONNX models on FPGAs (Sect. \ref{sec::proposed-flow}) featuring, for the first time to the best of our knowledge, both data-approximation and computation-approximation.
    \item The analysis of the effect of data-approximation in a mixed-precision tiny \ac{cnn} model for MNIST classification (Sect. \ref{ssec::exploration} and \ref{ssec::working-points}).
    \item The assessment of the benefits of computation-approximation through the deployment of an adaptive \ac{cnn} inference engine for the data-approximated models (Sect. \ref{ssec::adaptive}).
\end{itemize}

\section{Related Work} \label{sec::related}
\begin{figure}[htbp]
\centerline{\includegraphics[trim=3cm 5cm 2.5cm 4cm, clip, width=.9\linewidth]{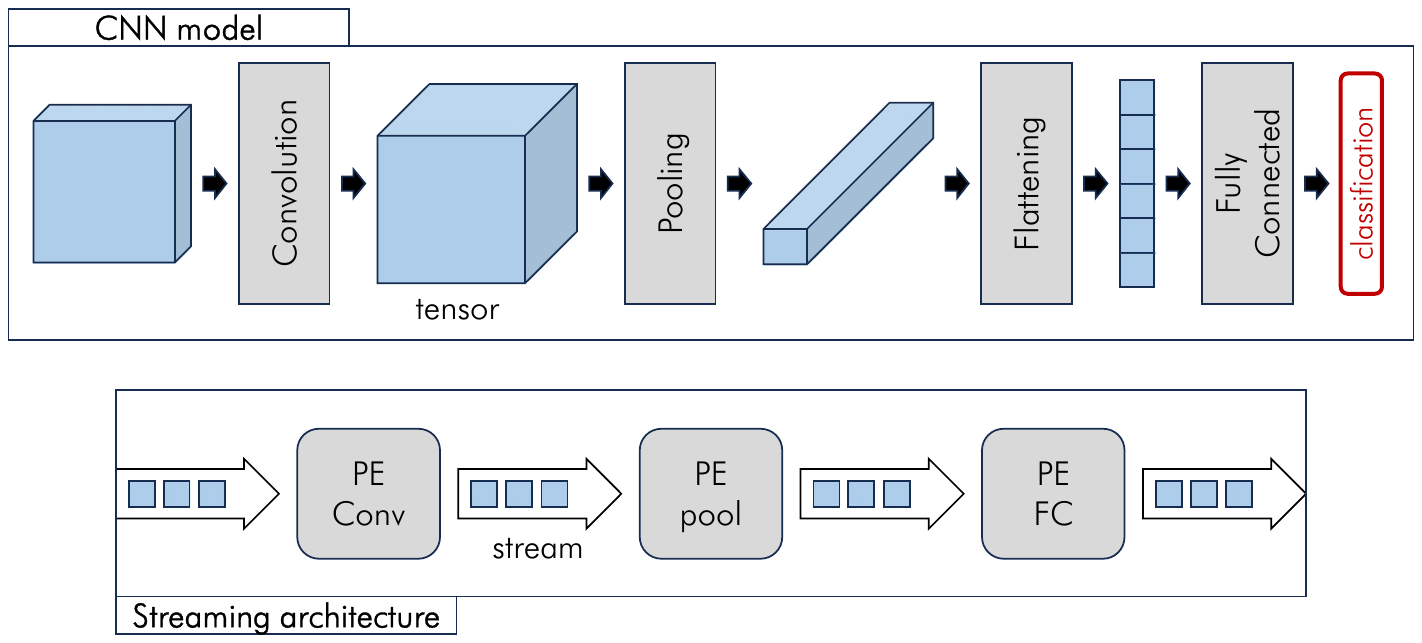}}
\caption{Representation on a simple CNN and its mapping to a streaming archietecture.}
\label{fig::streaming}
\end{figure}

To execute \ac{nn} at the edge, three main types of architectures can be found in literature~\cite{venieris2018}: the Single Computational Engine architecture, based on a single computation engine, typically in the form of a systolic array of processing elements or a matrix multiplication unit, that executes the CNN layers sequentially~\cite{systolic};
Vector Processor architecture, with instructions specific for accelerating operations related to convolutions~\cite{vector_processor}; 
the Streaming architecture consists of one distinct \ac{hw} block for each layer of the target CNN, where each block is optimized separately~\cite{aarestad2021,fraser2017}, as depicted in Fig. \ref{fig::streaming}. In this study, we adopted the latter for two main reasons: 
\begin{itemize}
    \item a distinct \ac{hw} processing element for each layer of the \ac{cnn} model allows for higher customization, thus favoring adaptivity;
    \item the streaming architecture is the most natural implementation of a dataflow-based application, such \acp{cnn}, thus easing the design with \ac{hls}.
\end{itemize} 

\subsection{Streaming Architectures} \label{ssec::streaming-archs}
In our previous work \cite{ratto2023}, a toolchain for porting \acp{cnn} on \acp{fpga} was proposed. The resulting \ac{hw} is a streaming architecture that uses on-chip memory, guaranteeing low-latency and low-energy computing.
Solutions that exploit a similar streaming architecture are FINN \cite{fraser2017}, a framework from AMD Research Labs; HLS4ML \cite{aarestad2021}, an open-source software designed to facilitate the deployment of machine learning 
models on FPGAs, targeting low-latency and low-power edge applications.
FINN enables building scalable and fast \ac{nn}, with a focus on the support of \ac{qnn} inference on
FPGAs. A given model, trained through Brevitas \cite{brevitas}, is compiled 
by the FINN compiler, producing a synthesizable C++ description of a heterogeneous streaming architecture. All
\ac{qnn} parameters are kept stored in the on-chip memory, which greatly reduces the power consumed and simplifies the design. The computing
engines communicate via the on-chip data stream. Avoiding the ``one-size-fits-all'', an ad-hoc topology is built for the network. The resulting accelerator is deployed on the target board using the AMD Pynq framework. 
The main operation of the HLS4ML library is to translate the model of the network into an HLS Project. The focus in \cite{diguglielmo2020} was centered on reducing the computational complexity and resource usage on a fully connected network for MNIST dataset classification: the data is fed to a multi-layer perceptron with an input layer of 784 nodes, three hidden layers with 128 nodes each, and an output layer with 10 nodes. 
The work exploits the potential of Pruning and \ac{qat} to reduce the model size with limited impact on its accuracy. 

\paragraph{This work positioning:}\label{sssec::hls4ml}To the best of our knowledge, neither FINN nor HLS4ML, despite targeting \ac{fpga}-based streaming architecture and supporting \ac{ac} features, have ever proposed an  adaptive solution. These existing frameworks primarily focus on data-oriented approximation. However, there remains an untapped opportunity for computation-oriented approaches, which can be achieved through reconfigurable systems design \cite{isca}. Such computation-oriented strategies could naturally be harnessed by runtime management infrastructures aiming at self-adaptive behaviors \cite{samos}, which are typical of \acp{cps}. 

\subsection{Approximate Computing}\label{ssec::ac}
The \ac{ac} paradigms is founded on \emph{``the idea that computer systems can let applications trade-off accuracy for efficiency''}.  Indeed,
\ac{ac} has been established as a design paradigm for energy-efficient circuits. It exploits the inherent ability of a large number of applications to produce results of acceptable quality despite being run on systems that \emph{``intentionally
exposes incorrectness to the application layer in return for conserving some
resource''\footnote{http://approximate.computer/approxbib/}}. 
This trade-off ultimately balances computation accuracy with efficiency. According to textbook definitions \cite{AgrawalCGGNOPSS16}, \ac{ac} provides three degrees of freedom by acting on \emph{data}, \emph{hardware}, and \emph{computation}. Approximating \emph{data} means processing either less up-to-date data (temporal decimation), less input data (spatial decimation), less accurate data (word-length optimization) or corrupted data. \emph{Hardware} approximation leverages inexact operators or voltage scaling. While \emph{computation} approximation corresponds to models modifications to expose different implementations, aiming to enable different execution profiles, over the same substrate. 

\Ac{ac} is particularly relevant in applications like \acp{nn} that have demonstrated remarkable resilience to errors \cite{mittal2016}. Within this specific application domain, \ac{nn} approximation can be broadly categorized into three main approaches: \emph{Computation Reduction}, \emph{Approximate Arithmetic Units}, and \emph{Precision Scaling}~\cite{ac_survey}. 
The \emph{Computation Reduction} approximation category aims at systematically
avoiding certain computation at the \ac{hw} level, thereby significantly reducing the overall workload. An example of this is pruning: biases, weights, or entire neurons can be evicted to lighten the workload~\cite{comp_reduction}.
By employing \emph{Approximate Units} that replace more accurate units, such as the Multiply-and-Accumulate (MAC) unit, energy consumption and latency in NN accelerators can be improved~\cite{approx_unit}. 
The most used \emph{Precision Scaling} practice is quantization:  quantized \ac{hw} implementations feature reduced bit-width dataflow and arithmetic units 
attaining substaintial energy, latency, and bandwidth gains compared to 32-bit floating-point
implementations. Instead of executing all the required mathematical operations with ordinary 32-bit floating point, quantization allows the exploitation of lighter operations by mapping real numbers to integers within a specified range \cite{precision_scaling}.

\paragraph{This work:} \ac{nn} \emph{Precision scaling} is exploited by implementing quantization to feature \emph{data} approximation. We combine different data-approximate profiles to enable \emph{computation} approximation and to deliver adaptivity. Our proposed flow utilizes Vitis HLS, which provides an arbitrary precision data types library, that goes beyond the standard C++. This library also supports customizable fixed-point data types\footnote{https://jiafulow.github.io/blog/2020/08/02/hls-arbitrary-precision-data-types}, easing the data precision control among layers. Additionally, we introduce another tool called MDC, explained further below, to enable adaptivity and \emph{computation} approximation.

\section{Proposed design flow} \label{sec::proposed-flow}
The utilization of a \ac{cnn} model involves two distinct phases: training and inference. The training phase aims at setting the model parameters to execute a given classification task. This phase tipycally occurs on powerful platforms, often in the cloud. The inference phase executes the trained model to perform the classification task. It is usually performed on a different platform, in our case an \ac{fpga} edge device. These two phases can be decoupled by adopting an intermediate representation to exchange the model between the training and the inference framework. The de facto standard for this purpose is the ONNX format.

The proposed design flow automates the design and deployment of an \ac{fpga} processor for the inference of a given Quantized \ac{cnn} model. The model must be provided in the QONNX format \cite{qonnx}, which extends the ONNX \footnote{https://onnx.ai/} format by allowing the specification of layers with arbitrary-precision data types. 
The adopted tools are described in Sect. \ref{ssec::tools} and their integration and usage in Sect. \ref{ssec::design-flow}.

\subsection{Tools} \label{ssec::tools}
Various commercial and open-source academic tools are utilized throughout the design flow:
\begin{itemize}
    \item the ONNXParser\footnote{https://gitlab.com/aloha.eu/onnxparser}, a Python application, is designed to parse the ONNX models and create the code for a target device. The tool consists of a Reader and multiple Writers, each tailored for different target platforms supported within the ALOHA framework \cite{aloha}. For this work, we developed a Writer targeting HLS.
    \item The Vitis HLS tool\footnote{https://www.AMD.com/support/documentation-navigation/design-hubs/dh0012-vivado-high-level-synthesis-hub.html} synthesizes a C or C++ function into RTL code for implementation on AMD FPGAs. The resulting \ac{hw} can be optimized and customized through the insertion of directives in the code.
    \item The \ac{mdc} tool\footnote{https://mdc-suite.github.io/} is an open-source tool that can offer Coarse-Grained reconfigurability support for \ac{hw} acceleration \cite{mdc}. It takes as input the applications specified as dataflow, together with the library of the HDL files of the actors. These dataflows are then combined, and the resulting multi-dataflow topology is filled with the actors taken from the HDL library. 
\end{itemize}

\subsection{Design Flow}\label{ssec::design-flow}
\begin{figure}[htbp!]
    \centering
    \hfill
    \subfigure
    {
    \includegraphics[trim=8.7cm 3.8cm 11cm 2.6cm, clip, width=.5\columnwidth]{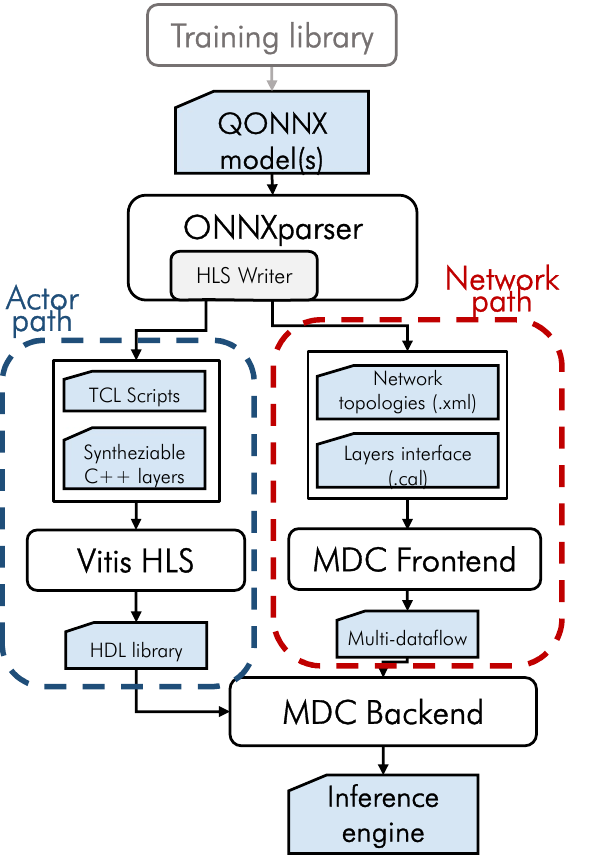}
    \label{aaa}
    } \hfill
    \subfigure
    {
   \includegraphics[trim=7.2cm 0cm 8.1cm 6cm, clip, width=.33\columnwidth]{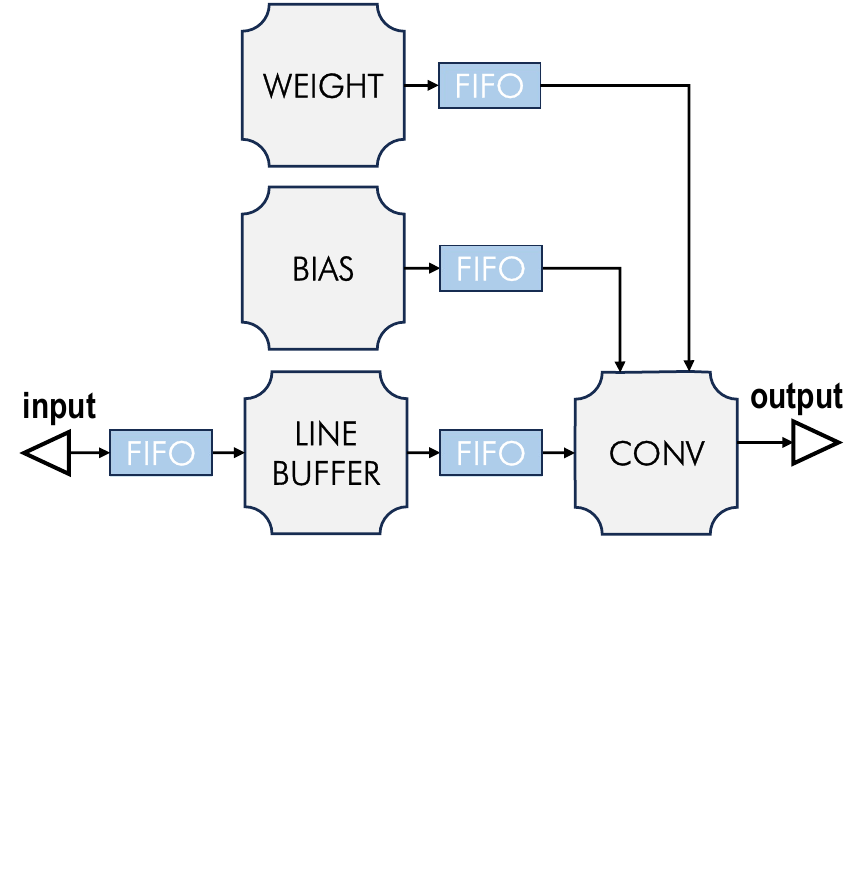}
    \label{bbb}
    }\hfill
    \caption{On the left, the ONNX-to-Hardware design flow for the generation of adaptive neural-network inference engines on FPGAs. The training library could be any library able to export to QONNX. On the right, the streaming-based template architecture for a convolutional layer.}
    \label{fig::design-flow}
\end{figure}

The proposed flow, as depicted in Fig.~\ref{fig::design-flow}, starts from the QONNX representation of the \ac{nn} and produces a streaming architecture that executes the input model. The QONNX file acts as a bridge between the training and the inference frameworks. Two distinct paths are present in the design flow: the actor related path and the network related path. They can be carried out once, to obtain a non-adaptive data-approximate solution, or multiple times, to derive a computation-approximate adaptive engine of data-approximate solutions.



The QONNX file serves as input to the extended ONNX Parser, which is capable of processing the additional quantization layers included in the QONNX format. Initially, the Reader reads the QONNX file and produces an intermediate format with a list of objects describing the layers' hyperparameters (e.g. kernel size, data precision, etc.) and connections within the QONNX model. Subsequently, the HLS Writer creates the target-dependent files. 

The ONNX Parser extracts the network topology from the QONNX and the data precision in each layer. This information is used by the Front End of the MDC tool to derive the datapath of the accelerator. When designing an adaptive engine, multiple data-approximate profiles of the same \ac{cnn} model are processed. The tool automates the merging process by sharing layers of different profiles that use the same data precision.

The HLS Writer produces the C++ files that implement the layers, and the TCL scripts to automate the synthesis by Vitis HLS. 
The C++ description of the layers is based on a template architecture: for the convolutional layers, the core of a CNN, the template is composed of a Line Buffer actor that stores the input stream to provide data reuse; the Convolutional actor, whose function is to execute the actual computation; and the Weight and Bias actors that store the kernel parameters needed for the convolution. The resulting template, depicted on the right side in Fig.~\ref{fig::design-flow}, ensures streaming dataflow between layers, eliminating the need to store full tensors. Each actor is developed to be customizable with the hyperparameter, e.g. input and kernel size, extracted from the QONNX model. The HDL library produced by Vitis HLS and the reconfigurable datapath (Multi-datflow) serve as input to the MDC Backend to generate the HDL description of the inference engine.


\section{Evaluation} \label{sec::evaluation}
The proposed design flow was evaluated on a tiny CNN model trained for MNIST classification. The model comprises two convolutional blocks and a final fully connected layer. Each block consists of a convolutional layer with a 3x3 kernel, 64 filters, and ReLU activation, followed by a batch normalization and a maxpooling layer. 
The inference engines that execute this model have been designed with the proposed flow targeting the FPGA available on an AMD KRIA board.

First, we describe how the quantized models have been trained in Sect. \ref{ssec::training}. Then, to evaluate the proposed flow, we carried out an initial exploration, described in Sect. \ref{ssec::exploration}, on \emph{data} approximated designs. This exploration is meaningful to assess the impact of quantization on both accuracy and inference performance. Then, in Sect. \ref{ssec::working-points}, different execution profiles are selected and, then, merged to generate an adaptive inference engine, as described in  \ref{ssec::adaptive}.

\subsection{Quantization-aware training} \label{ssec::training}
The model previously described has been designed and trained using QKeras \cite{qkeras}. QKeras is an extension of the Keras framework that offers several features, including the ability to specify a custom fixed-point precision for each layer of a NN and perform \ac{qat}, , which has demonstrated significant advantages over post-training quantization in terms of resulting model accuracy \cite{gholami2021}

Through its APIs, QKeras allows to specify the number of bits used to represent the activations and the weights of the NN model. The activations are the outputs of an NN layer, while the weights are the trainable parameters of the kernel used either for convolution, in convolutional layers, or matrix multiplication in fully connected layers. In the exploration of Sect. \ref{ssec::exploration} we varied both these bit-widths, thus implementing a mixed-precision quantization strategy. 

For the \ac{qat}, we selected an optimizer that implements the \emph{Adam algorithm} and \emph{Categorical Crossentropy} as the loss function to be minimized during regression. The trained \ac{cnn} model have been exported to QONNX format and implemented with the proposed design flow. 

It is worth underlying that other frameworks, e.g. Brevitas \cite{brevitas}, offer \ac{qat} and export to QONNX, so that the proposed design flow is interoperable with any QONNX-compliant framework.

\subsection{Data approximation analysis} \label{ssec::exploration}
In this section, we report the results on the analysis using mixed-precision quantization. A string identifies a profile as \textit{Ax-Wy}, where \textit{x} represents the number of bits used to represent activations and \textit{y} the number of bits used for weights. For each mixed-precision configuration, a non-adaptive inference engine has been realized with the proposed design flow . We report accuracy, latency for a classification, resource utilization, and power consumption for each engine in Table~\ref{tab::results}. 

\begin{table*}[htb]
\renewcommand\arraystretch{1.0} 
\caption{Results of the analysis with data mixed-precision approximation. In the string \textit{Ax-Wy}, \textit{x} represents the number of bits for the activations, and \textit{y} is the number of bits used for the weights.}
	\centering
\begin{tabular}{l c c c c c}
\toprule
\multirow{2}{*}{Datatype}   &Accuracy    &Latency  &LUT    &BRAM     &Power\\
       &[\%]  &[us]  &[\%]  &[\%] &[mW]\\ \midrule
A16-W8  &98.9       &329       &12       &18       &160   \\ 
A16-W4  &95.3       &329       &7       &18       &134   \\ 
A8-W8  &98.8       &329       &11       &17       &142   \\ 
A8-W4  &95.3       &329       &6       &17       &132   \\ 
A4-W4  &95.8       &329       &6       &17       &141   \\  
\bottomrule
\end{tabular}
\label{tab::results}
\end{table*}

It can be noticed that the execution latency for an image classification remains constant independently of the data precision. This behavior can be explained considering how the HLS compilation flow works: the HLS compiler schedules the operations depending on data dependencies and user directives. After that, the operations are bound to the physical resources. Therefor, larger bit precision increases computing resource utilization rather than slowing down the system.  Indeed, we can see that adopting a reduced bit precision for activations and weights leads to a reduction in \acp{lut} and \acp{bram} utilization.

The two metrics where we see an exploitable trade-off at runtime are model accuracy and power consumption. The model's accuracy decreases with reduced bit precisions. From a baseline 99.8\% which can be obtained with floating point operations, not feasible to be ported to an FPGA, the quantized model A16-W8 achieves a classification accuracy of 98.9\%. This accuracy drops down to  95.3\% with A8-W4. We can notice that with 4-bit precision in the weights, the final accuracy is around 95\%. Small variations are due to the intrinsic randomness of the training process rather than to the activations' precision. 

On the other hand, this drop in accuracy is compensated by reduced dynamic power consumption. A general trend shows that power consumption decreases with reduced precision. A graphical description of the resulting execution profiles that consider only accuracy and power consumption is reported in Fig. \ref{fig::pareto}. 
The variability in the power consumption, which is not directly proportional to the data precision, shows the advantage of having a fast design flow that goes from the high-level description in QONNX to the FPGA implementation. This allows us to consider the joint effects of the resource utilization, which is affected by the FPGA backend and the HLS compiler, and switching activity, which depends on the actual values of weights and the data being processed.

\subsection{Execution profiles selection} \label{ssec::working-points}
\begin{figure}[htbp]
\centering
\includegraphics[trim=1cm 3.3cm 1cm 2.5cm, clip, width=.6\columnwidth]{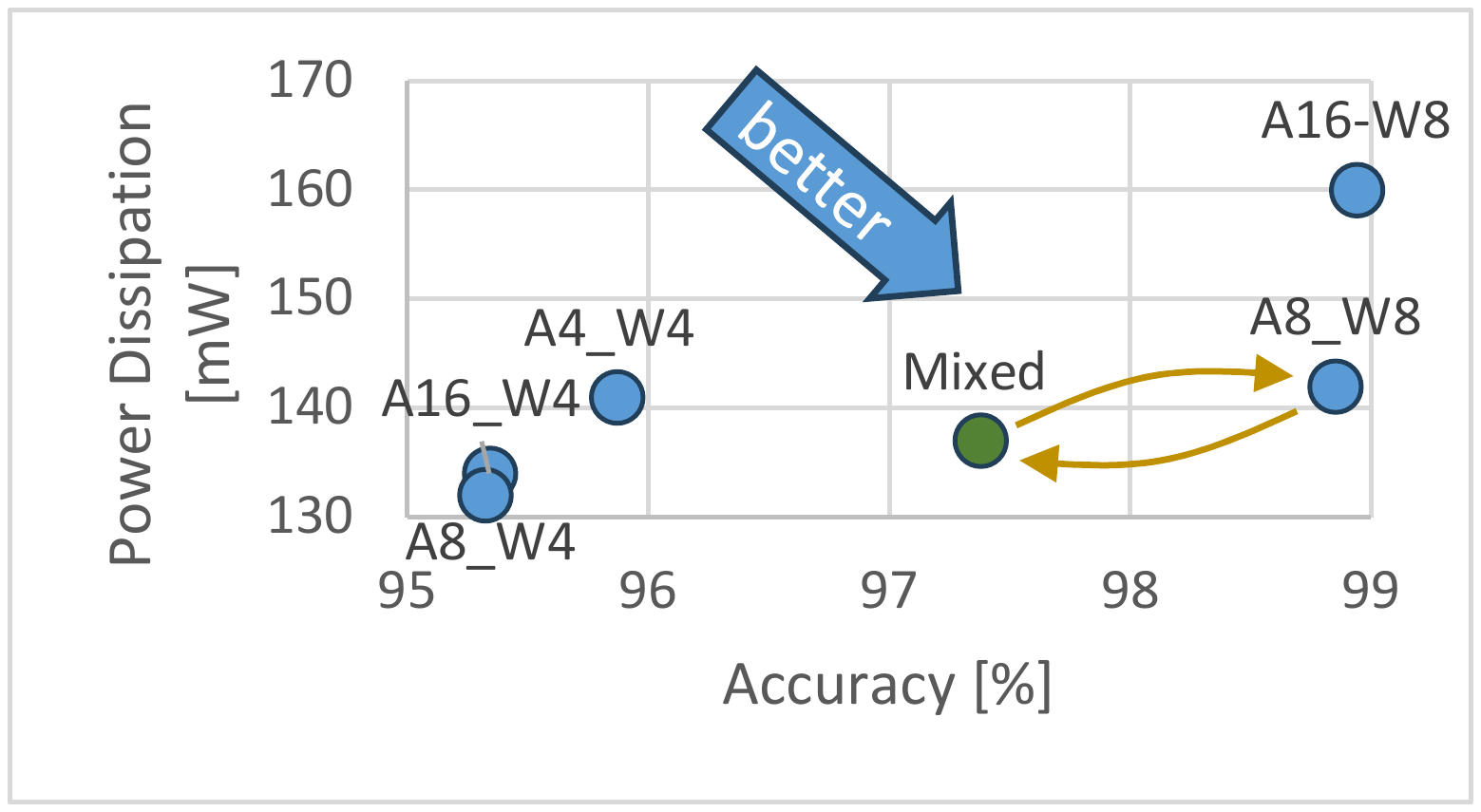}
\caption{Accuracy VS power chart of the obtained profiles. In green the Mixed design. The yellow arrows point to the two configurations selected for adaptivity.} 
\label{fig::pareto}
\end{figure}

From Fig. \ref{fig::pareto}, we observe that the non-adaptive inference engines obtained with the initial exploration offer valuable trade-offs. However, these engines lack common layers necessary to achieve some degree of resource sharing. To address this limitation, we started from the A8-W8 profile and trained an additional profile that further exploits mixed precision. This new profile generally uses the same precision as A8-W8, but in the inner convolutional layer, where instead it uses the A4-W4 one. The resulting non-adaptive engine (named Mixed) performance is reported with a green dot in Fig. \ref{fig::pareto}. This demonstrates the additional level of \emph{data} approximation that can be achieved with the proposed methodology. 

Finally, the Mixed and A8-W8 profiles are good candidates for merging using the proposed methodology. This will allow us to design an adaptive inference engine that enables  \emph{computation} approximation, as shown in the following section.

\subsection{Adaptive inference engine} \label{ssec::adaptive}
\begin{figure}[htbp]
\centerline{\includegraphics[trim=4.5cm 6.7cm 4.5cm 6.2cm, clip, width=\columnwidth]{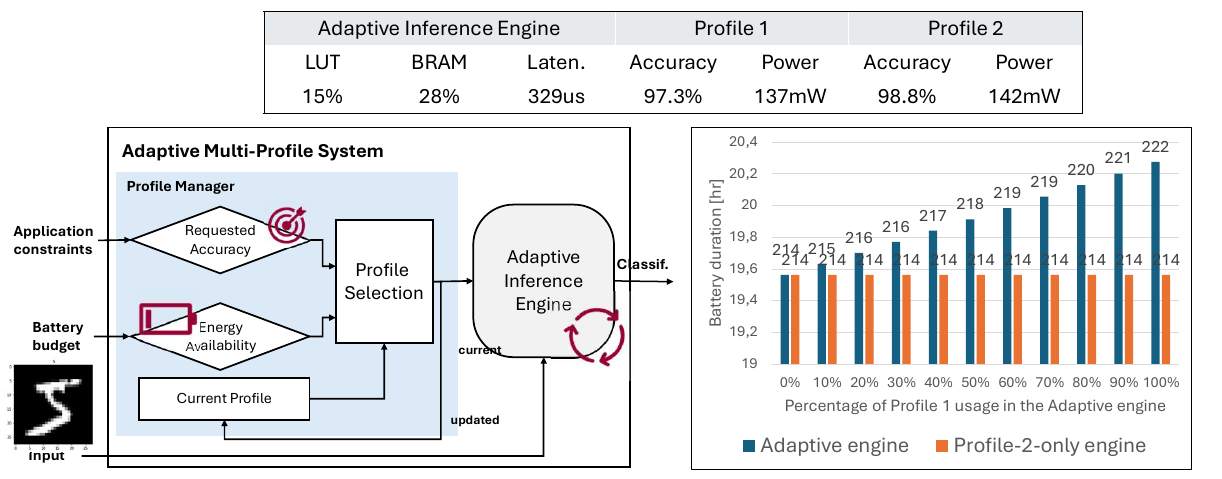}}
\caption{On top the resource utilization of the adaptive engine and its perfomance metrics under different profiles. On the left side the architecture of a complete adaptable systems that exploits the proposed adaptive inference engine. On the right side, a comparison of the resulting battery duration (supposing a 10Ah energy budget) and number of classifications executable by the adaptive engine and a non-adaptive one supporting the higher accuracy profile only.} 
\label{fig::battery}
\end{figure}

In the previous designs we partially used the functionality of the proposed methodology, resulting in non-adaptive inference engines. To achieve adaptivity we need to design a \emph{computation} approximate inference engine that allows selecting different profiles at runtime. For this purpose, we leverage the merging capabilities of MDC. As anticipated, Mixed and  A8-W8 profiles are selected as entry points, since they share the same layers, but the inner convolutional one. The characteristics of the resulting adaptive inference engine are summarized on top of Fig. \ref{fig::battery}. The resulting inference engine has a limited overhead with respect to the non-adaptive ones. The switch among profiles can guarantee a 5\% power saving with a 1.5\% accuracy drop. Given the low accuracy penalty, we can suppose that in a real CPS application, the inference engine would run most of the time in the Profile 1 and switch to the more accurate only under critical circumstances, when higher accuracy is necessary. This further motivates the proposed methodology that is going to be adopted as part of a recently started EU project\cite{cf24}. 

Indeed, a \ac{cps} is meant to react and dynamically adjust to mutable constraints and system conditions. This can be achieved, as shown on the left-hand side in Fig. \ref{fig::battery}, by an infrastructure composed of two main parts: the \emph{Adaptive Inference Engine} and the \emph{Profile Manager}. The former is responsible of implementing the adaptive solution that, in this case, can alternatively execute one of the two profiles. The latter, following the self-adaptive management approach presented in \cite{samos}, monitors the energy status and the given constraints and decides which is the most suitable profile. The profile selected at runtime must capable of meeting the accuracy requirements while minimizing power dissipation. As an example, if the remaining battery budget is lower than a pre-defined threshold the \emph{Profile Manager} might select a less energy consuming profile, if the user/application defined constraints are still met or if they can be negotiated. 
On the right-hand side in Fig. \ref{fig::battery}, the potentials of the implemented adaptive engine are presented. Even considering this preliminary implementation, it is shown how the adaptive engine (in blue) extends the battery duration, and in turn increases the number of executable classifications, with respect to the non-adaptive (in orange) counterpart, which is running at full performance. 

\section{Conclusion} \label{sec::conclusion}
CPS integrate computation with physical processes, characterized by information exchange with the environment and dynamic behaviors. FPGAs offer \ac{hw} acceleration, flexibility, and energy efficiency, but challenges persist in achieving full adaptivity for dealing with complex environments.

The utilization of a CNN model involves two distinct phases: training and inference. These phases can be decoupled using an intermediate representation like the ONNX format. The proposed design flow automates FPGA inference for quantized CNN models, specified in the QONNX format, which allows data approximation through arbitrary-precision data types. At the same time, the flow also features adaptivity, implementing computation approximation. 

A data approximation analysis on a tiny CNN model for MNIST classification has been carried out to select valuable execution profiles. These latter have been automatically combined in a runtime adaptive inference engine, which is capable of adapting its accuracy and power consumption at runtime by switching among the selected profiles.

Future work will aim at validating the proposed approach on more complex CNN models and datasets, allowing for quantitative state of the art comparison, besides the already provided qualitative discussion.

\emph{\footnotesize{This work is supported by MYRTUS that is funded by the European Union (GA No. 101135183). Views and opinions expressed are however those of the author(s) only and do not necessarily reflect those of the European Union. Neither the European Union nor the granting authority can be held responsible for them.}}


\begin{thebibliography}{8}

\bibitem{aarestad2021}
Aarrestad, Thea, et al. "Fast convolutional neural networks on FPGAs with hls4ml." Machine Learning: Science and Technology 2.4 (2021).

\bibitem{AgrawalCGGNOPSS16} Agrawal, Ankur  et al. "Approximate computing: Challenges and opportunities." ICRC Conference 2016.

\bibitem{ac_survey}
Armeniakos, Giorgos, et al. "Hardware approximate techniques for deep neural network accelerators: A survey." ACM CSUR 55.4: 1-36 (2022).

\bibitem{approx_unit}
Bhardwaj, Kartikeya, et al. "Power-and area-efficient approximate wallace tree multiplier for error-resilient systems." ISQED symposium 2014.

\bibitem{Legup}
Canis, Andrew, et al. "LegUp: An open-source high-level synthesis tool for FPGA-based processor/accelerator systems." ACM TECS 13.2 (2013): 1-27.



\bibitem{qkeras}
Coelho, Claudionor N., et al. "Automatic heterogeneous quantization of deep neural networks for low-latency inference on the edge for particle detectors." Nature Machine Intelligence 3.8:675-686 (2021).

\bibitem{vector_processor}Farabet C., et al.
"An FPGA-based processor for convolutional networks."
FPL Conference 2009.

\bibitem{fraser2017}
Fraser, Nicholas J., et al. "Scaling binarized neural networks on reconfigurable logic." PARMA-DITAM Workshop 2017.

\bibitem{gholami2021}
Gholami, Amir, et al. "A survey of quantization methods for efficient neural network inference." Low-Power Computer Vision Book (2021).

\bibitem{systolic}
Guan, Yijin, et al. "FP-DNN: An automated framework for mapping deep neural networks onto FPGAs with RTL-HLS hybrid templates." FCCM Symposium 2017.

\bibitem{guo2019}
Guo, Kaiyuan, et al. "[DL] A survey of FPGA-based neural network inference accelerators." ACM TRETS 12.1: 1-26 (2019).

\bibitem{precision_scaling}
Jungwook, Choi, et al. "Accurate and Efficient 2-bit Quantized Neural Networks." MLSys Proceedings
2019.


\bibitem{aloha}
Meloni, Paolo  et al. "Optimization and deployment of CNNs at the edge: the ALOHA experience." CF Conference 2019.

\bibitem{mittal2016}
Mittal, Sparsh. "A survey of techniques for approximate computing." ACM CSUR 48.4: 1-33 (2016).

\bibitem{isca} Nezan, Jean-Francois et al. Multi-purpose systems: A novel dataflow-based generation and mapping strategy." ISCAS Symposium 2012.

\bibitem{diguglielmo2020}
Ngadiuba, Jennifer, et al. "Compressing deep neural networks on FPGAs to binary and ternary precision with hls4ml." Machine Learning: Science and Technology 2.1 (2020).

\bibitem{samos} Palumbo, Francesca et al. "Hardware/Software Self-adaptation in CPS: The CERBERO Project Approach." SAMOS Conference 2019.

\bibitem{cf24}
Palumbo, Francesca et al. "MYRTUS: Multi-layer 360° dYnamic orchestration and
interopeRable design environmenT for compute-continUum
Systems." CF Conference 2024.

\bibitem{brevitas}
Pappalardo, Alessandro, "Xilinx/brevitas". Zenodo (2023), https://doi.org/10.5281/zenodo.3333552

\bibitem{qonnx}
Pappalardo, Alessandro, et al. "Qonnx: Representing arbitrary-precision quantized neural networks." AccML Workshop 2022.

\bibitem{ratto2023}
Ratto, Francesco, et al. "An Automated Design Flow for Adaptive Neural Network Hardware Accelerators." Journal of Signal Processing Systems (2023): 1-23.


\bibitem{mdc}
Sau, Carlo, et al. "The Multi-Dataflow Composer tool: An open-source tool suite for optimized coarse-grain reconfigurable hardware accelerators and platform design." MICPRO Journal 80 (2021).

\bibitem{edge_ai}
Shafique, Muhammad, et al. "TinyML: Current progress, research challenges, and future roadmap." DAC Conference 2021.

\bibitem{comp_reduction}
Song Han, et al. "EIE: Efficient Inference Engine on Compressed Deep Neural Network." ACM SIGARCH Computer Architecture News 44.3 (2016).

\bibitem{maxcompiler}
Summers, Sioni, et al. "Using MaxCompiler for the high level synthesis of trigger algorithms." Journal of Instrumentation 12.02 (2017).


\bibitem{umuroglu2017}
Umuroglu, Yaman, et al. "Finn: A framework for fast, scalable binarized neural network inference." FPGA Symposium 2017.

\bibitem{venieris2018}
Venieris, Stylianos, et al. "Toolflows for mapping convolutional neural networks on FPGAs: A survey and future directions." ACM CSUR 51.3: 1-39 (2018).




\end{thebibliography}
\end{document}